\documentstyle[eqsecnum,twocolumn,aps]{revtex}
\begin{document}
\draft
\title{
Dynamic Response of Ising System to a Pulsed Field}
\author{M. Acharyya$^{1,2,*}$, J. K. Bhattacharjee$^3$ 
and B. K. Chakrabarti$^4$}
\address{ $^1$Department of Physics, 
 Indian Institute of Science, Bangalore-560012, India \\ 
 $^2$Condensed Matter Theory Unit,
 Jawaharlal Nehru Centre for Advanced Scientific Research,
 Jakkur, Bangalore-560064, India \\
$^3$ Indian Association for the Cultivation of Science,
Jadavpur, Calcutta-700 032, India \\
$^4$ Saha Institute of Nuclear Physics,
1/AF Bidhannagar, Calcutta-700 064, India.}

\maketitle
\narrowtext
\begin{abstract}
The dynamical response to a pulsed
magnetic field has been studied here both using Monte Carlo 
simulation and by solving numerically the
meanfield dynamical equation of motion 
for the Ising model. The ratio $R_p$ of the response magnetisation half-width
to the width of the external field pulse
has been
observed to diverge 
and pulse susceptibility $\chi_p$ (ratio of the response magnetisation peak 
height and the pulse height) gives a peak near the order-disorder 
transition temperature $T_c$ (for the unperturbed system).
The Monte Carlo results for Ising system on square lattice show that $R_p$
diverges at $T_c$,
with the exponent $\nu z \cong 2.0$, while $\chi_p$ shows a peak at $T_c^e$,
which is a function of the field pulse width $\delta t$. A finite size (in
time) scaling analysis shows that $T_c^e = T_c + C (\delta t)^{-1/x}$,
with $x = \nu z \cong 2.0$. 
The meanfield results show that both the divergence of $R$
and the peak in $\chi_p$ occur at the meanfield transition temperature,
while the peak height in $\chi_p \sim (\delta t)^y$, $y \cong 1$ for 
small values of $\delta t$. These results also compare well with an approximate
analytical solution of the meanfield equation of motion.
\end{abstract}


\section{Introduction}
The dynamic response of the Ising systems has recently been studied 
extensively employing computer simulations [1]. In particular, the study of
dynamical response of Ising systems to oscillating magnetic field [2,3] has
led to many intriguing dynamic phenomena, like dynamic hysteresis and the
fluctuation induced dynamic symmetry breaking transitions in (low e.g., 
one, two
or three dimensional) Ising system in presence of an oscillating field.
Acharyya and Chakrabarti also noted [3,4] some
anomalous behaviour in the growth of pulse 
susceptibility, in Ising systems under
pulsed magnetic fields of finite durations. 

Usually, when a cooperatively interacting thermodynamic system in 
equilibrium is perturbed (with the perturbation having a step function like
variation with time), then the relaxation of the system (to the equilibrium
appropiate to the perturbed state) is observed to follow
the common Debye type form
with a single relaxation time. The standard (Debye) form for any response
function (say magnetisation of Ising system) $m(t)$ is
\begin{eqnarray}
m(t) \sim m(\infty) + A ~{\rm exp} (-t/\tau),
\end{eqnarray}
\noindent where $\tau$ is the relaxation time, $m(\infty)$ denotes the new
equilibrium value, $A$ is a constant. As the critical temperature is
approached $\tau$ shows a 
critical slowing down; $\tau$ diverges at the critical
temperature $T_c$:
\begin{eqnarray}
\tau \sim \xi^z \sim (T - T_c)^{-\nu z},
\end{eqnarray}
\noindent where $\xi$ is the correlation length, $z$ is the dynamic exponent
and $\nu$ is the correlation length exponent [1].

Here, we have investigated in details the response of pure Ising systems to
pulsed magnetic fields of finite duration, using Monte Carlo simulations for
two dimensional Ising systems, and solving numerically the meanfield equation 
of motion. 
We have studied the response behaviour for 'positive' pulses, where the
pulsed field is in the direction of the sponteneous magnetisation in
the ordered phase. In the disordered phase, of course, this notion is
immaterial. One can also study the effect of 'negative' pulses on the
sponteneous order, where the field direction is opposite to sponteneous
magnetisation. Although many intriguing features of the domain growth etc
are expected for such 'negative' pulse problem, we confine here to the study
of 'positive' pulses only.
We have measured the ratio $R_p$ 
of the response magnetisation (pulse)
half-width ($\Delta t$) 
to that ($\delta t$) of the pulsed external field, and the ratio $\chi_p$ of the
response magnetisation peak height ($m_p$) to the field pulse height 
($h_p$) giving the
pulse susceptibility. The temperature variation of these two quantities for 
various pulse width durations and heights of the external field, have been
investigated.

We find that for weak pulses, while the width-ratio $R_p$ diverges at the 
order-disorder transition point $T_c$ of the unperturbed Ising system,
the pulse susceptibility $\chi_p$ does not diverge and for low dimensions,
e.g., in two dimension studied here, a smeared peak in $\chi_p$ occurs at an
effective $T_c^e$, which approaches $T_c$ as the field pulse width $\delta t$
increases ($\chi_p$ diverges at $T_c$ as $\delta t \to \infty$). A meanfield
analysis for the absence of divergence of $\chi_p$ for finite $\delta t$ has
been developed. Also, a finite time scaling analysis, similar to Fisher's
finite size (length) scaling [5], has been developed and compared with the
observation of the effective transition temperature $T_c^e$ with the
external field pulse width $\delta t$.

We have organised this paper 
as follows: In section II, the model 
and the simulation techniques have been 
described.  In section III, the results are given. The paper ends with  
concluding remarks in section IV.

\section{The model and simulation}

A ferromagnetically interacting 
(nearest neighbour) Ising system in presence of
a time dependent magnetic field can be described by the Hamiltonian
\begin{equation}
H = -\sum_{ij} s^z_i s^z_j - h(t) \sum_i s^z_i,
\label{ham}
\end{equation} 
where $s^z_i$'s are spin variable having their value $\pm 1$ 
and $h(t)$
is the time varying longitudinal magnetic field. Here, we have considered the
time variation of $h(t)$ as follows:
\begin{eqnarray}
h(t) & = & h_p ~~~{\rm for}~~~t_0 < t < t_0 + \delta t \nonumber \\
& = & 0 ~~~{\rm elsewhere,} \label{fld}
\end{eqnarray}
where $h_p$ is the amplitude of the field and $\delta t$ is the
duration or the active period of the external field.

In our simulation, we have considered a 500$\times$500 square lattice in two
dimension.
At
each site of the lattice there is a spin variable $s^z_i$ = ($\pm 1$). We 
update the lattice by stepping sequentially over it following the
Glauber single spin flip dynamics. One such full scan over the lattice
is unit time step (Monte Carlo step or MCS)
here. First, we allowed the system to reach the equilibrium (at any
temperature $T$) and only after that the
magnetic field $h(t)$ has been switched on ($t_0$ is thus much larger than the
relaxation time of the system). 
One can also use random updating sequences. However, it takes more time (MCS)
to stabilise the system. We expect all the features of the response studied
here remains qualitatively unchanged for random updating sequences, and we
give the results here for sequential updating only.
We have measured the maximum height (above the
equilibrium value) $m_p$ and half-width $\Delta t$ of the response 
magnetisation. Here, as mentioned before, 
the following two quantities have been defined to
characterise the response of the system:
$${\rm pulse ~~width ~~ratio} ~~R_p = {{\Delta t} \over {\delta t}}$$
\noindent and
$${\rm pulse ~~susceptibility} ~~ \chi_p = {{m_p} \over {h_p}}.$$
It may be noted that $\chi_p$ reduces to normal static susceptibility
as $h_p \to 0$ and $\delta t \to \infty$. At each temperature for fixed $h_p$
and $\delta t$,
the numerical values of $R_p$ and $\chi_p$ are obtained averaging over 
20 random Monte
Carlo realisations (initial seed).
We have studied the temperature variation of these two 
quantities. The results are given and discussed in the next section. The 
observation for finite peak in $\chi_p$ at an effective transition
temperature $T_c^e ~~({\rm which ~~converges ~~to}~~ 
T_c ~~{\rm as}~~\delta t \to \infty)$ 
is analysed in
view of finite time scaling behaviour, as mentioned before.

To study the similar response in the case where the fluctuations are absent,
we have considered the following meanfield
dynamical equation of motion for the kinetic Ising system:
\begin{equation}
\tau_0 {dm \over dt} = -m + tanh\left({{m + h(t)} \over T} \right). 
\label{mfe} 
\end{equation}
 Here $\tau_0$ is the microscopic relaxation time, 
 $m$ is the average magnetisation (in the meanfield approximation), 
$h(t)$ is the time varying pulse field having the same time variation 
described in equation \ref{ham} and $T$ denotes the temperature. 
We have solved numerically the above equations using fourth order Runge-Kutta
method.
We have 
evaluated the pulse width ratio $R_p$ and pulse susceptibility $\chi_p$ at
various
temperatures for fixed pulse width $\delta t$ and height $h_p$.  
The numerical results are given in the next section, where we have compared the
results for finite peak height in $\chi_p$ (at $T_c$ = 1), with an approximate
analytic estimate of $\chi_p$ in such cases.

\section{Results}
\subsection{Monte Carlo studies}

As mentioned before, our results here are for two dimensional Ising system 
on square lattice of size 500$\times$500.
We applied an external field of amplitude $h_p$ for a duration 
 $\delta t$ after bringing the system into a steady state. For this, the typical
 number of Monte Carlo steps required for this size of the lattice chosen here,
 is observed to be of the order of $10^6$.
 The response magnetisation has amplitude $m_p$ (measured
from the equilibrium value) and a half-width $\Delta t$. 
Fig. 1 shows a typical time variation of magnetic field $h(t)$ and the 
corresponding response magnetisation $m(t)$.
The dynamical
response is characterised by two quantities; the width-ratio $R_p$ and the
pulse-susceptibility $\chi_p$. The temperature variation of these two
quantities has been studied. Fig. 2 shows the temperature 
variations of $R_p$
and $\chi_p$ for  fixed values of $h_p  (= 0.5)$ and for three values of
$\delta t ~~(= 5, 10, 25$ MCS).

Since $m_p$ is bounded from above, a large $h_p$ would saturate 
$m_p$ and hence $\chi_p$ becomes small (due to the saturation). Also, for
extremely small $h_p$, it becomes difficult to identify $m_p$ from the
noise, and hence the estimate of $\chi_p$ becomes erroneous. We found
$h_p \cong 0.5$ to be well within the above optimal range.
From the figure it is 
clear that $R_p$ has
a sharp divergence at $T_R  (\cong 2.30,
~~{\rm somewhat ~~larger ~~than}~~  T_c$, the Onsager transition
temperature, due to the small size
of the system)
 almost irrespective of the values of the pulse width $\delta t$.
But $\chi_p$ shows peak  
at different points 
(significantly above $T_c \sim 2.27$)
depending upon the values of the field pulse 
width $\delta t$.
As $\delta t$ value increases, it is observed that the peak shifts towards
$T_c$ from above (and also the peak height grows). 

Let us try to understand why the width-ratio $R_p$
diverges at $T_c$, while height-ratio or pulse susceptibility
$\chi_p$ shows peak at some
higher value $T_c^e(\delta t)$ depending upon the value 
of $\delta t$: $T_c^e(\delta t) \to T_c$ as $\delta t \to \infty$. 
The pulse-like perturbation probes the response of the system at finite 
frequencies.
Consequently, the $\chi_p (=m_p/h_p)$ can not diverge as 
the response magnetisation height $m_p$ is not an 
 equilibrium value corresponding
to the pulse height $h_p$; rather the $m_p$
results are
bounded by the time window of width $\delta t$. On the other
hand, the response will take its own relaxation time to come to its 
equilibrium value (irrespective of the value of $\delta t$),
when the field is switched off (at $t_0 + \delta t$). 
This leads to the divergence of $\Delta t$, due to critical slowing
down, as $T$ approaches  $T_c$. 

The sharp divergence of width-ratio $R_p$ is identified as the consequence of
critical slowing down and the point of divergence is the critical temperature
$T_c$ for the ferro-para transition. 
In fact, since the relaxation after the withdrawal of the pulse
is unrestricted by the pulse-width, we can
assume that $\Delta t \sim \tau \sim |T - T_c|^{-\nu z}$. We therefore plot
in Fig. 3,  $R_p^{-1/\nu z}$ versus $T$ and find a straight line plot with
$\nu z \cong 2.0$ in two dimensional case. This compares well with the previous
estimates of the value of $\nu z$ [6].

As the growth of the height of magnetisation response (and its maximum 
value $m_p$) is very much bounded by the time window $\delta t$ of the applied
field pulse, the anomalous behaviour of $\chi_p$ (having finite peak at a
shifted temperature $T_c^e(\delta t)$) may be considered to be due to the finite
size (in time) effect. Similar to the
finite size (in length) scaling theory of Fisher [5], where a finite size 
system shows effective (non-singular or non-divergent)
pseudo-critical behaviour at $T_c^e(L)$ when the
correlation length $\xi$ becomes of the order of the system size $L$, we
suggest a finite time $(\delta t)$ scaling behaviour here for $\chi_p$: If
the relaxation time $\tau \sim \xi^z \sim |T-T_c|^{-\nu z}$, where $\nu$ is the
correlation length exponent and $z$ is the dynamic exponent, then $\chi_p$
would show peak at the temperature $T_c^e$ here when 
$\tau(T_c^e) \sim \delta t$ or 
$|T_c^e(\delta t)-T_c|^{-\nu z}
\sim \delta t$, 
 or 
\begin{equation}
T_c^e(\delta t) \sim T_c + C (\delta t)^{-1/\nu z}, \label{fsh}
\end{equation}

\noindent  where $C$ is some constant. In fact, Fig. 4 shows that 
the effective peak 
 position $T_c^e$ indeed approaches $T_c$ as $\delta t \to \infty$. The inset
 shows the plot of $T_c^e$ with $(\delta t)^{-1/x}$, which gives a
 straight line for $x = \nu z \cong 2.0$. This again suggests $\nu z \cong 2.0$
 and also the extrapolated value of $T_c$ becomes 
about 2.29, which
 compares well with the Onsager value, comparable to the
 previous estimate. 

\subsection{Meanfield results}

 We have solved the meanfield equation for response magnetisation \ref{mfe} 
using fourth
order Runge-Kutta method. Fig. 1 shows the typical variation of response
magnetisation and field. Here also we have measured the width-ratio and
the pulse susceptibility and studied the temperature variation of these two
quantities.

Fig.5 shows the temperature variation of $R_p$ and $\chi_p$ for different
values of $\delta t$. $R_p$ diverges and $\chi_p$ peaks
at the same order-disorder transition point ($T_c = 1$ here). 
We have also studied the variation of the maximum value of $\chi_p$ 
($\chi_p^{max}$ at $T = T_c$) with respect to the 
duration of the pulsed field.
It may also be mentioned that the
 peak height was found to increase with 
increasing pulse width $\delta t$, and $\chi_p$ 
 diverges as $\delta t \to \infty$: $\chi_p^{max} 
\sim (\delta t)^y; y \cong 1.0$
Fig. 6 shows the variation of $\chi^{max}_p$ with $\delta t$. 

In order to comprehend these observations, we solve the equation \ref{mfe} in a 
linearised limit (large $T$ and small $h_p$; specifically $T > 1$, $h_p \to 0$).
 In such limit, the equation of motion becomes
\begin{equation}
\tau_0 {dm \over dt} = -\epsilon m + h(t)/T; ~~\epsilon = (T-1)/T. \label{lin} 
\end{equation}
One can solve the above equation using a form
$m(t) = m_0 e^{-t/\tau}$ which gives
\begin{eqnarray}
\tau_0 {dm_0 \over dt} e^{-t/\tau} - {\tau_0\over \tau} m_0 e^{-t/\tau} 
= -\epsilon m_0 e^{-t/\tau} + {h(t)\over T},
\end{eqnarray}
giving $\tau/\tau_0 = \epsilon^{-1}$, and

\begin{equation}
\tau_0 {dm_0 \over dt}e^{-t/\tau} = h(t)/T. \label{lin1}
\end{equation}
Integrating the last equation for $h(t) = h_p$ for a finite time 
width $\delta t$ and $h(t) = 0$ elsewhere, one gets 
$m_p \sim h_p \delta t/(T_c \tau_0)$
at $T = T_c = 1 ~~({\rm when~~} \tau \to \infty)$. This gives

\begin{equation}
\chi_p^{max} = \chi_p(T_c)= m_p(T_c)/h_p \sim \delta t/(T_c \tau_0) 
\sim \delta t/\tau_0. \label{rp}
\end{equation}
 We have checked the above linear relationship of $\chi_p^{max}$
 with $\delta t$ for extremely small values of pulse field
amplitude $h_p$ (see Fig. 6).

\section{Summary}

We have studied the Glauber (order parameter nonconserving) dynamics
of an Ising system under a time varying external magnetic field, when
the field is applied as a pulse of finite time width, after the system
reaches the equilibrium. The time variation of the response magnetisation
is studied as a function of pulse width $\delta t$, height $h_p$ and
temperature $T$ of the system. We have measured specifically
$R_p = {\Delta t}/{\delta t}$ and $\chi_p = m_p/h_p$, where $\Delta t$ is
the time width of the response magnetisation and $m_p$ is the maximum
height of the response magnetisation above its equilibrium value.

Our computer simulation results for square lattice showed
$R_p \sim |T_c - T|^{-\nu z}$ with $T_c \cong 2.30$, the Onsagar value,
and $\chi_p$ has a peak $\chi_p^{max}$ at $T_c^e > T_c$, such that a
finite size (in time) scaling behaviour is observed: $T_c^e = T_c
+ C (\delta t)^{-1/{\nu z}}$, with $\nu z \cong 2.0$. The numerical
solutions of the mean field equation \ref{mfe} showed $R_p \sim 1/(T-1)$ and
the pulse susceptibility peak value $\chi_p^{max} \sim \delta t$,
occuring at $T = T_c = 1$. Theoretical analysis for the finite size
(in time) scaling behaviour (of $T_c^e$ in the Monte Carlo case) and
for the peak $\chi_p^{max} (\delta t)$ (in the mean field case) are
also given.

\section*{Acknowledgements}

MA acknowledges JNCASR for financial support and SERC, IISc Bangalore
for computational facilities.

\centerline {\bf Figure Captions}

\noindent Fig.1. Time variation of magnetic field ($h(t)$) and the response
magnetisation ($m(t)$)
in the Monte Carlo case. For $h_p$ = 0.5 and $\delta t$ = 50.

\noindent Fig.2. Temperature variations of 
(a) $R_p$ and (b) $\chi_p$ for different
values of $\delta t$
in the Monte Carlo case. The symbol
circle is for $\delta t$ = 5; the square is for $\delta t$ = 10
and the cross is for $\delta t$ = 25.

\noindent Fig.3. Variation of $R_p^{-1/{\nu z}}$ against $T$, with 
$\nu z$ = 2.

\noindent Fig.4 Variation of $T_c^e$ with respect to $1/{\delta t}$
in the Monte Carlo case. Inset shows
the variation of $T_c^e$ with respect to $(\delta t)^{-1/x}$; $x$ = 2.

\noindent Fig.5 Temperature variation of (a) $R_p$ and 
(b) $\chi_p$ for different
values of $\delta t$ in the mean field case. 
The symbol triangle is for $\delta t$ = 8;
the plus is for $\delta t$ = 16 and the cross is for $\delta t$ = 32.

\noindent Fig.6. Variation of $\chi_p^{max}$ with respect to $\delta t$ in the
mean field case.
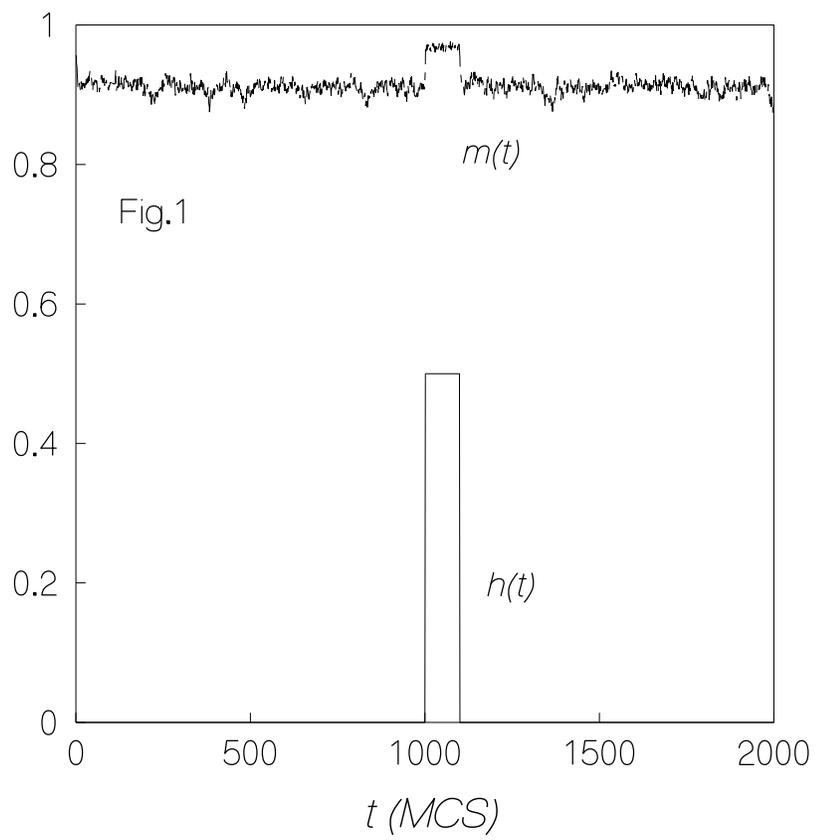
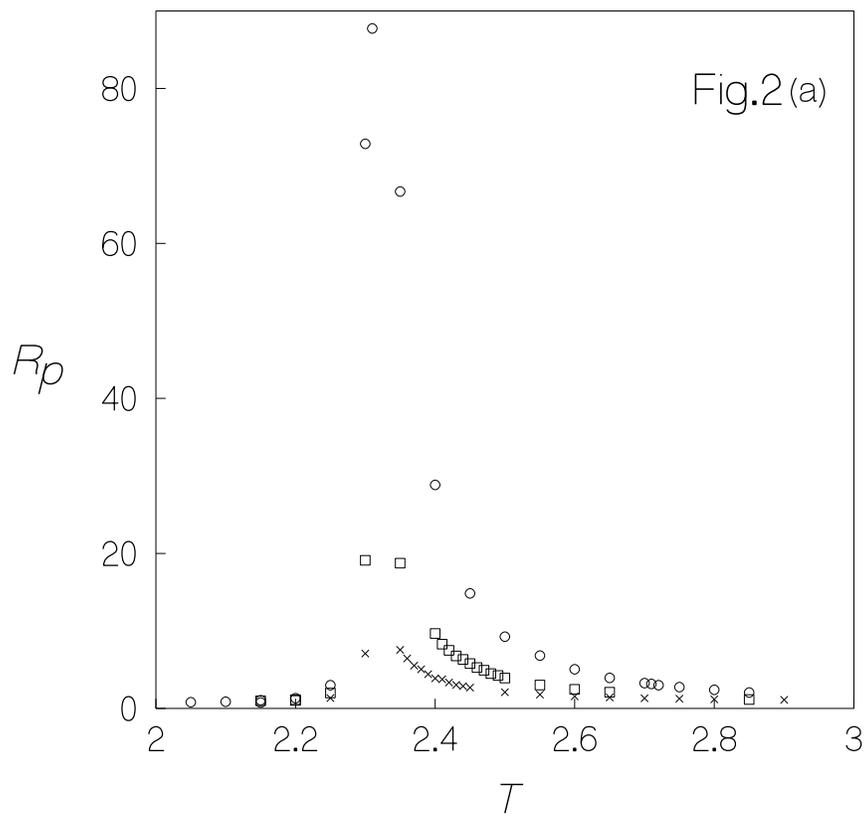
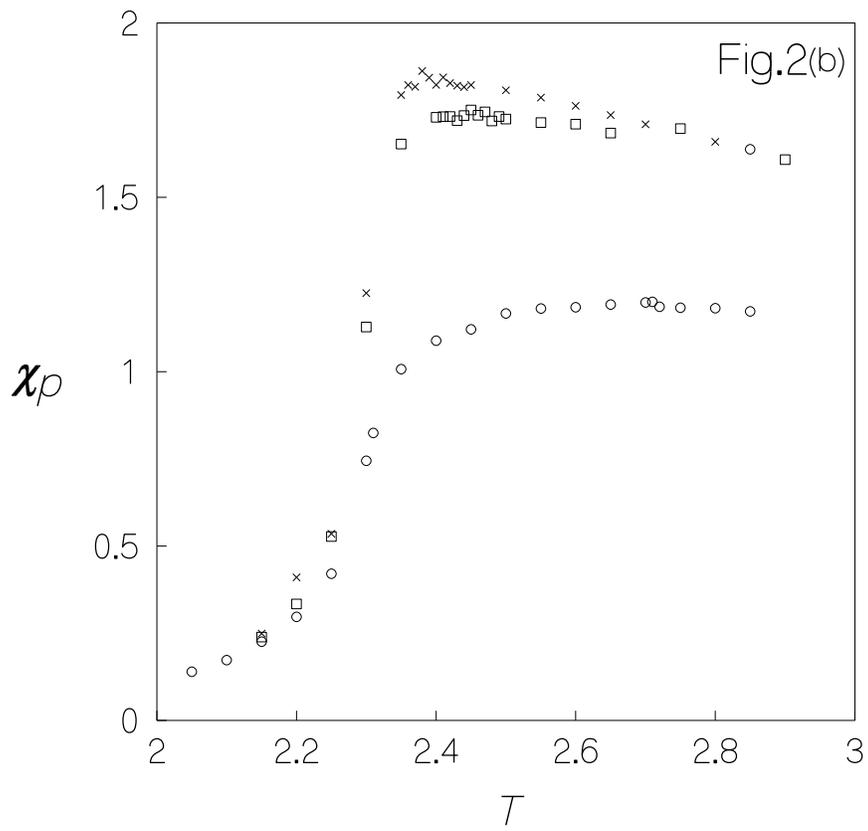
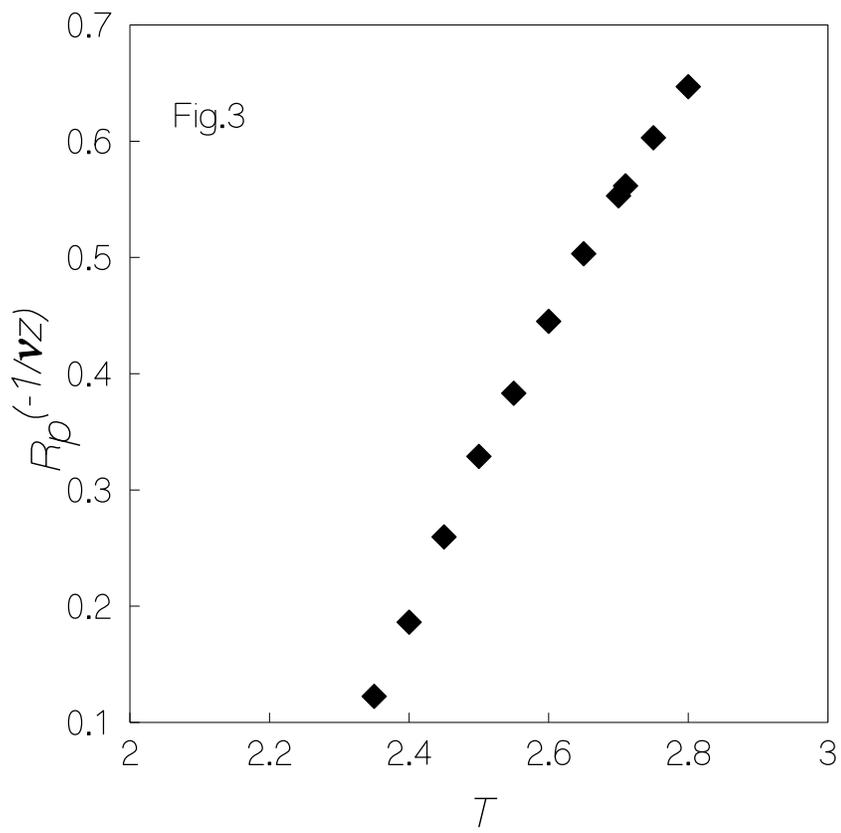
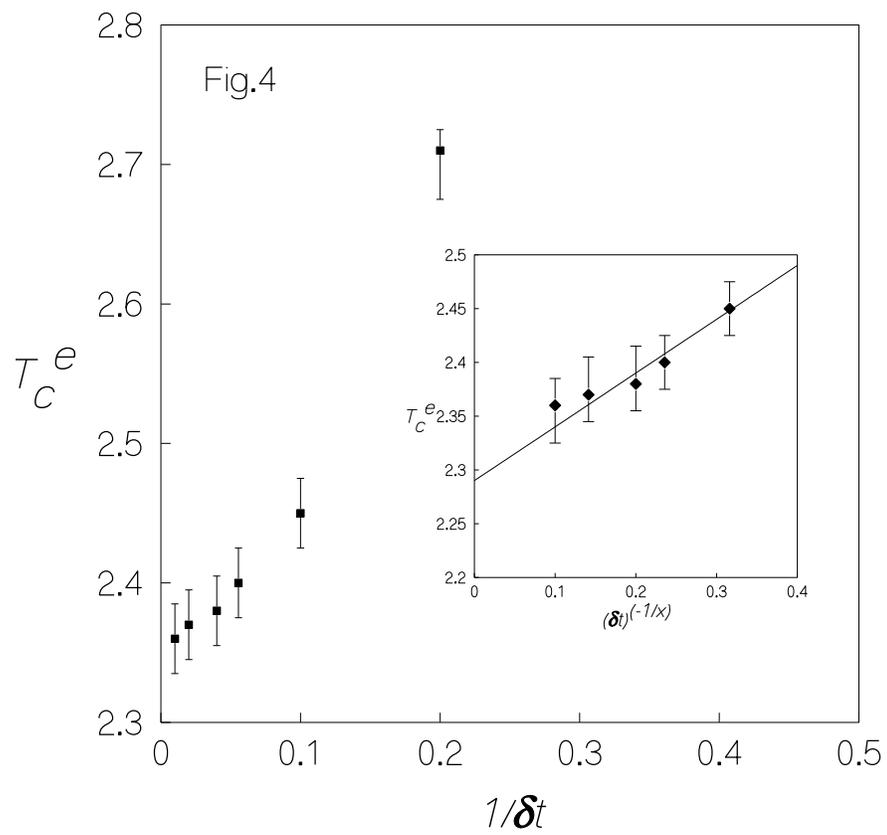
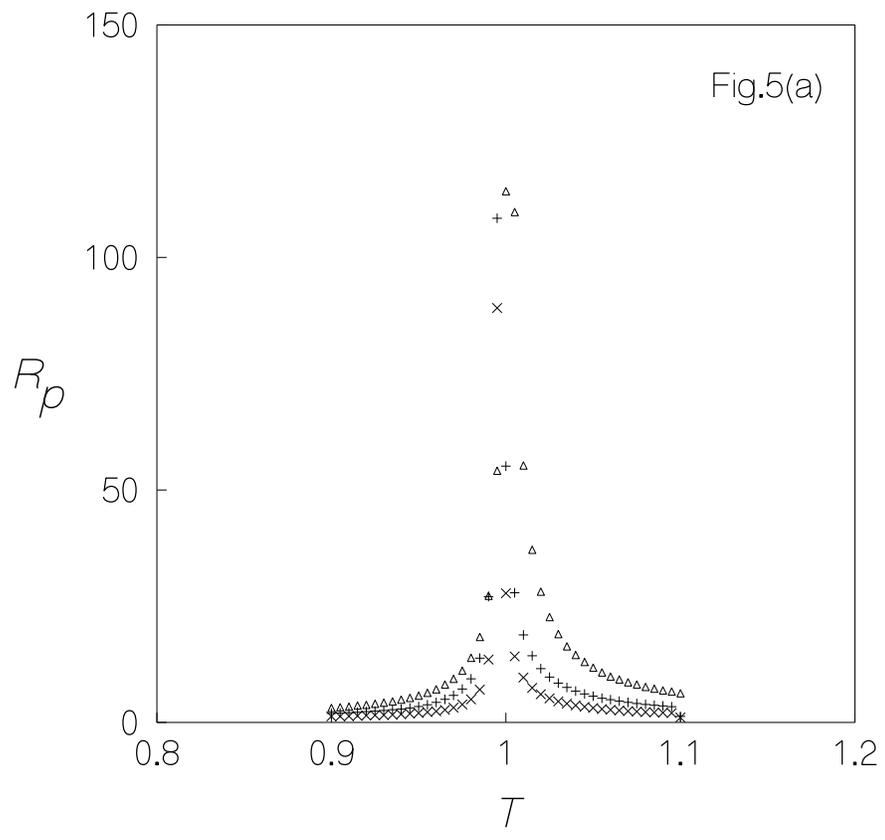
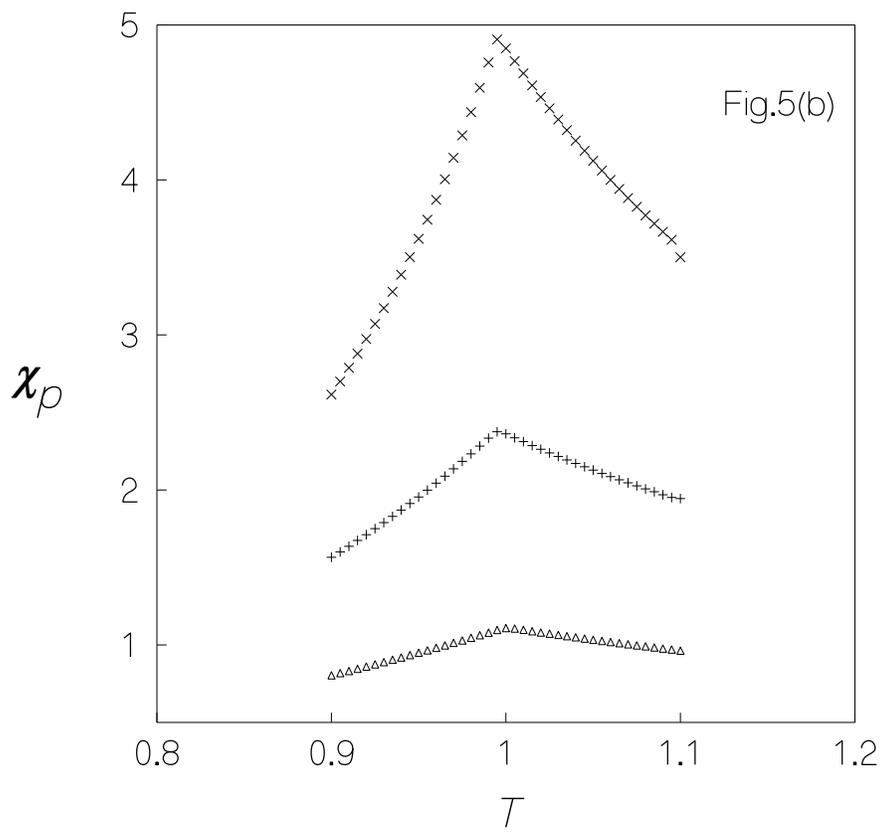
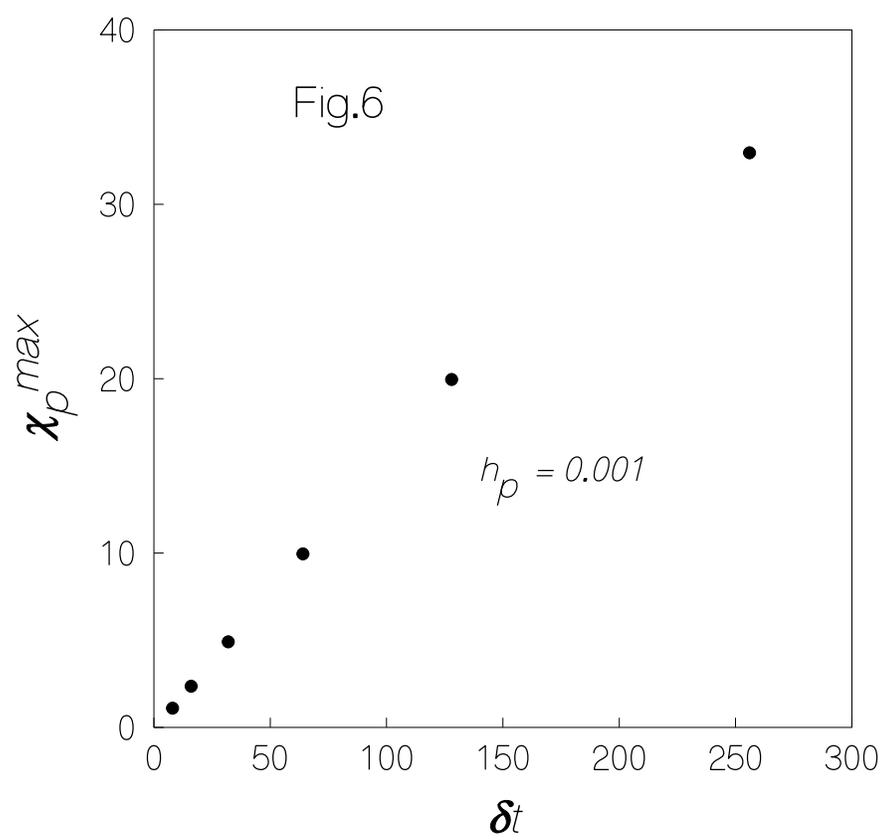
\end{document}